\documentclass[a4paper,fleqn,usenatbib,useAMS]{mnras}
\usepackage[T1]{fontenc}
\usepackage{aecompl}
\usepackage{graphicx}	
\usepackage{amsmath}	
\usepackage{amssymb}

\usepackage{graphicx,color}

\title{Magnetic Rayleigh-Taylor Instability in Radiative Flows}

\author[ A. Yaghoobi \& M. Shadmehri]{Asiyeh Yaghoobi $^{1}$, Mohsen Shadmehri$^{2}$\thanks{E-mail:
m.shadmehri@gu.ac.ir; mmshadmehri@gmail.com}\\
$^{1}$ Department of Physics, Institute for Advanced Studies in Basic Sciences (IASBS), PO Box 11365-9161, Zanjan, Iran \\
$^{2}$ Department of Physics, Faculty of Sciences, Golestan University, Gorgan 49138-15739, Iran }

\begin{document}

\maketitle

\date{Received ______________ / Accepted _________________ }

\label{firstpage}

\begin{abstract}
We present a linear analysis of the radiative  Rayleigh-Taylor (RT) instability in the presence of magnetic field for both optically thin and thick regimes. When the flow is optically thin, magnetic field not only stabilizes perturbations with short wavelengths, but also growth rate of the instability at long wavelengths is reduced compared to a nonmagnetized case. Then, we extend our analysis to the optically thick flows with a conserved total specific entropy and properties of the unstable perturbations are investigated in detail. Growth rate of the instability at short wavelengths is suppressed due to the presence of the magnetic field, however, growth rate is nearly constant at long wavelengths  because of the radiation field. Since  the radiative bubbles around massive protostars are subject to the RT instability, we also explore implications of our results in this context. In the nonmagnetized case, the growth time-scale of the instability for a typical  bubble is found less than one thousand years which is very short compared to the typical  star formation time-scale. Magnetic field with a reasonable strength significantly increases the growth time-scale to more than hundreds of thousands years. The  instability, furthermore, is more efficient at  large wavelengths, whereas in the non-magnetized case, growth rate at short wavelengths is more significant. 
\end{abstract}

\section{Introduction}
 Different types of  instability have been proposed to understand some features of the astrophysical objects or formation of the structures in the  astrophysical flows. Rayleigh-Taylor (RT) instability  is an important mechanism which operates at the interface of two fluids with different densities while accelerating towards each other \citep{chandra}. The RT instability has found applications in various astrophysical systems, such as expansion of the supernova remnants \citep[e.g.,][]{ribeyre},  bubbles in the intracluster medium \citep[e.g.,][]{pizzolato,jiang,krum12},  prominences  in the solar atmosphere \citep[][]{terradas}, and, interior of the red giants \citep[e.g.,][]{charbonnel}.

Linear growth rate of the RT instability and its nonlinear evolution have been studies by many authors. For an incompressible flow, \cite{chandra} showed that the RT instability is suppressed, if the surface tension and the viscosity are considered.  The effect of compressibility on the RT instability has been studied by \cite{shiva} who showed that  growth rate of the short-wavelength unstable perturbations decreases due to the compressibility. Since the RT instability may has an efficient role in some of the magnetized astrophysical systems, the classical incompressible RT instability has also been extended to the magnetized case for a configuration with a uniform magnetic field parallel to the interface of the flows \citep{chandra}. The linear analysis shows that a tangential magnetic field reduces  growth rate of the RT instability. Subsequent studies extended magnetized RT instability to the partially ionized systems where ions and neutrals as separate fluids can exchange momentum through direct collisions \citep[e.g.,][]{diaz12,shadmehri13,diaz14}. It is shown ion-neutral collision reduces the linear growth rate of the RT instability. Another important physical factor is the radiation field which may have a dynamically significant role in the RT instability. The radiation force, for instance, is significant at the boundary of a radiation-driven H II region \citep[e.g.,][]{Jacquet}, or during massive star formation \citep[e.g.,][]{Krumholz09,kumar,kuiper}. Radiative RT instability has been studied by a few authors subject to  simplifying assumptions \citep[e.g.,][]{Mathews77,Krolik77}. Detailed radiative RT instability in the linear regime for optically thin and thick systems has been formally analyzed by \cite{Jacquet}  (hereafter JK) for configurations with an isotropic radiation pressure. In the optically thin and isothermal regime, they showed that role of the radiation field is like an effective gravitational field, and,  asymmetry of the  H II regions can be attributed to the RT instability. In the optically thick and  adiabatic regime, the RT growth rate is determined at the long-wavelength limit, however, it tends to a finite value when the radiation is close to the Eddington limit (JK). In short-wavelength limit the effect of the radiation field is negligible on the growth rate of the RT instability. Role of radiation field on the RT instability for a background state with a pure scattering opacity has been studied by \cite{jiang} in both linear and nonlinear regimes. They solved  the radiation hydrodynamic equations numerically with anisotropic radiation pressure in interface. The obtained growth rate of the RT instability exhibits a reduction in the presence of the radiation field and it decreases with radiation pressure. Results of \cite{jiang} also showed that anisotropy of radiation plays an important in the nonlinear development of the RT instability.

Regarding to the importance of the magnetic fields, we extend the analysis of JK to the magnetized case with a uniform tangential magnetic field  at both the upper and the lower flows. Basic equations are presented in section 2. We then investigate RT instability in the optically thin flows in section 3. A generalized dispersion relation including magnetic field and radiation is obtained. In section 3, our analysis of the magnetic RT instability is extended to the optically thick flows.  We then conclude with astrophysical implications of our results in section 5. 
\section{PLASMA CONFIGURATION AND EQUATIONS}
Basic equations of the Radiative Magnetohydrodynamic (RMHD) have been introduced and discussed by \cite{Stone} which are written in a commoving frame and are accurate up to $O(u/c)$, where $u$ is the fluid velocity and $c$ is the light speed. The momentum and the energy are exchanged between the material and the radiation. A similar set of the RMHD equations have also been implemented by \cite{Lowrie} and \cite{Blaes} in the equilibrium diffusion approximation. Upon neglecting gas self-gravity and assuming the flow is subject to a uniform gravity, the main RMHD equations are
\begin{equation}\label{eqs1}
\frac{D\rho}{Dt}+\rho \nabla\cdot{\bf u}=0,
\end{equation}
\begin{equation}\label{eqs2}
\rho\frac{D{\bf u}}{Dt}=-\nabla p_{\rm g}+\rho{\bf g}+\frac{1}{4\pi}(\nabla\times{\bf B})\times{\bf B} + \frac{\kappa_{\rm F} \rho }{c} {\bf F},
\end{equation}
\begin{equation}\label{eqs3}
\rho \frac{D }{Dt}(\frac{E_r}{\rho})=-\nabla \cdot {\bf F} -\nabla {\bf u} : {\cal P}_r  -4\pi\rho {\kappa_p} {\textit{B}_p}-c\rho\kappa_E E_r ,
\end{equation}

\begin{equation}\label{eqs4}
\rho \frac{D }{Dt}(\frac{E_g+E_r}{\rho})=-\nabla \cdot {\bf F} -\nabla {\bf u} : {\cal P}_r  -  p_{\rm g} \nabla\cdot \bf{u} ,
\end{equation}
\begin{equation}\label{eqs5}
\frac{\rho}{c^2}\frac{D }{Dt}(\frac{{\bf F}}{\rho}) + \nabla \cdot {\cal P}_{\rm r}=-\frac{\kappa_{\rm F}\rho}{c} {\bf F},
\end{equation}
\begin{equation}\label{eqs6}
 \frac{\partial{\bf B} }{\partial t}=\nabla\times({\bf u}\times{\bf B}).
\end{equation}
{Here, the Lagrangien derivative is denoted by $D/Dt \equiv \partial /\partial t + {\bf u}.\nabla $ and ${\bf u}$ is the velocity of the flow and $ E_r, F,{\cal P}_r$ and $E_g$ , are the comoving radiation energy density, flux and stress tensor and the internal energy density, respectively and $ {B}_p $ is Planck function. The gas density, pressure and internal energy per unit volume are denoted by $\rho$, $p_g = \rho c_s^2$, and $E_g = p_g/(\gamma -1)$, respectively. Furthermore, $\kappa_F$, $\kappa_p$ and $\kappa_E$ are the flux mean, Planck mean and energy mean opacities. Moreover, the isothermal sound speed is $c_s=\sqrt{k_{B}T/m}$, where $k_B$, $T$ and $m$ are the Boltzman constant, the temperature and the mean mass per particle, respectively.  The flow is assumed to be in the local thermal equilibrium. The above equations are closed with suitable relations for the equation of state, opacities and the Planck function and also a tensor variable Eddington factor $f$,  i. e. $ {\cal P}_r=fE_r$ (\cite{Stone}). In the following sections, we simplify the above basic equations under various approximations. We note that equation (\ref{eqs4}) can also be re-written as
\begin{equation}\label{eqs7}
\frac{DE_{\rm t}}{Dt}+{\cal H}_{\rm t} : \nabla {\bf u}=-\nabla \cdot {\bf F}.
\end{equation}
This equation is actually equation (10) of JK. Here, we have  $E_t = E_r + E_g $, ${\cal H}_t = E_t {\cal I}_3 + {\cal P}_t$, ${\cal P}_t = p_g {\cal I}_3 + {\cal P}_r$ and ${\cal I}_3$ is the $3\times 3$ identity matrix. 
\begin{figure}\
\includegraphics[width=8cm]{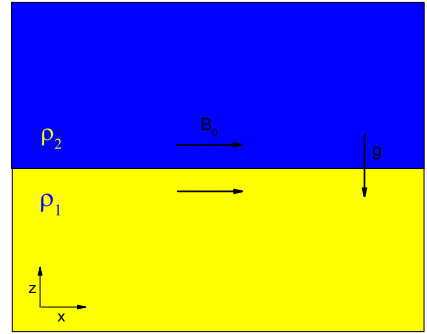}
\caption{ A fluid with the density $\rho_2$ is superposed  on a lighter one, with density $\rho_1$, such that the gravitational acceleration $\bf g$ is perpendicular to the interface. An initial uniform magnetic field in parallel to the interface exists at both the upper and the lower layers.}
\label{f1}
\end{figure}
\\

The adopted geometry of our model is shown in Figure \ref{f1}. In a Cartesian coordinates, we assume that the plane $z=0$ specifies the interface between two fluids with different densities. We denote the 
physical quantities of the lower fluid $(z < 0)$ by a subscript 1, and, those in the upper fluid  $(z> 0)$ are marked with a subscript 2. We assume that the initial magnetic field is uniform and tangent to the interface, i.e. $\textbf{B} = B_0\widehat{x}$, whereas the 
gravitational acceleration is perpendicular to the interface, i.e. $ \textbf{g}= -g\widehat{z}$.
\section{ANALYSIS IN THE OPTICALLY THIN REGIME}
We study the RT instability in an isothermal and optically thin   system. We thereby suppose that both the upper and the lower fluids are kept at a fixed temperature and  the radiation flux is assumed to be constant. Possible astrophysical implication of the present study are  ionization fronts around H II regions (e.g., JK), the quasi-stellar object (QSO) clouds that may experience an intense central radiation force  \citep[e.g.,][]{mathews}, and the radiation driven outflows in the ultraluminous infrared galaxies \citep{jiang}.\\

 The main equations  for this regime are written as
\begin{equation}\label{otin1}
\frac{D\rho}{Dt}+\rho \nabla\cdot{\bf u}=0,
\end{equation}
\begin{equation}\label{otin2}
\rho\frac{D{\bf u}}{Dt}=-\nabla p_{\rm g}+\rho{\bf g}+\frac{1}{4\pi}(\nabla\times{\bf B})\times{\bf B} + \frac{\kappa_{\rm F} \rho }{c} {\bf F},
\end{equation}
\begin{equation}\label{otin3}
 \frac{\partial{\bf B} }{\partial t}=\nabla\times({\bf u}\times{\bf B}).
\end{equation}
This set of equations is closed with the ideal gas equation of state, i.e. $p=k_{B}T\rho/m$.
\subsection{Initial equilibrium state}

We assume that the initial velocity of the flow is zero, and, the gas density and the radiation pressure are functions of the vertical coordinate $z$. Vertical component of the equation of motion leads to the following relation:
 \begin{equation}\label{otin4}
 \frac{\partial p_g }{\partial z}=-\rho  g+\frac{\kappa_{F} \rho }{c} F= -\rho {g_{\rm eff}}.
 \end{equation}
where we suppose material with pure scattering opacity in both sides of the interface. The specific scattering $\kappa_F$ and $\bf{F}$ are assumed to be constant \citep[JK,][]{jiang}. So an effective gravitational acceleration is defined as $ \bf{g_{\rm eff}}= \bf{g} $ $(1-E)$ which is a constant, but may differ in the upper and lower layers, because the radiation flux may be different in the upper and lower layers. Here, the parameter $E$ denotes the Eddington limit of the background state and is written as $E=({\kappa_{\rm F}}/{c} F)/({g})$. The above equation, therefore, can also be rewritten as
  \begin{equation}\label{otin5}
 \frac{\partial p_g }{\partial z}=-\rho  g(1-E).
 \end{equation}

In the optically thin regime, we consider only electron scattering following previous studies \citep[JK,][]{jiang}. In our work, the Eddington  parameter $E$ is not dependent on the other parameters like $\alpha$ and $\beta$. We neglect role of radiation by setting $E=0$. This particular case can be interpreted as a configuration in which either the opacity or radiation flux of one layer is much smaller than other layer. This approximation has already been implemented by JK in analyzing HII region around bright stars.  In the optically thick regime,  we can assume the opacity is a constant when the adiabatic approximation is used (JK).  We note that when the gravitational force and the force of radiation are in balance, the Eddington limit is the maximum luminosity a body (such as a star) can achieve while keeping the system in equilibrium. For example, if radiation of a star exceeds the Eddington luminosity, the surface layers are no longer in equilibrium and one can expect a very intense radiation-driven stellar wind from its outer layers. As we mentioned earlier,  the initial magnetic field $B_0$ is  assumed to be uniform and tangent to the interface.  The initial magnetic field is the same at both upper and lower layers because of the continuity of the magnetic pressure. But the density and the temperature are different at two sides of the interface.
\subsection{Linear perturbations}
Using the above initial equilibrium state, we can now perturb each physical quantity as $\chi\equiv\chi_0+\chi'$, where the perturbed quantity is much smaller than the equilibrium state, i.e.  $\vert \chi'\vert\ll\vert \chi_0\vert $. Upon substituting linear perturbations into equations \ref{otin1}, \ref{otin2}, \ref{otin3} and the equation of state, a set of linear differential equations is obtained such that one can study their time-evolution as  $\chi' (z,x,t)=\chi'(z)\exp(\omega t+ik_x x)$, where the primed quantities are the amplitude of the perturbations, $\omega$ is the growth rate of the instability, and, $k_x$ is wavenumber of perturbations. We note that with this form of time-dependence part of the perturbations, real values of $\omega$ imply exponential growth which then correspond to the unstable modes.  We then arrive to the following linearized differential equations:
\begin{equation}
 \omega\rho'+u'_z\frac{\partial \rho }{\partial z}+\rho (ik_x u'_x+\frac{\partial u'_z}{\partial z}) =0, 
\end{equation}
\begin{equation}
\rho  \omega u'_x=-ik_x p', 
\end{equation}
\begin{equation}
 \rho  \omega u'_z=-\frac{\partial p'}{\partial z}-\rho' g_{\rm eff}+\frac{B_0^2}{4\pi\omega}(\frac{\partial^2 u'_z}{\partial^2 z}-k^2_x u'_z), \\
\end{equation}
\begin{equation}
 p'=c_s^2\rho' .\label{otin6}
\end{equation}
Note that equilibrium state is denoted by quantities without subscript, for simplicity. After straightforward mathematical manipulations, we obtain
\begin{equation}\label{otin7}
\frac{\partial^2 u'_z}{\partial z^2}-\frac{g_{\rm eff}}{c_s^2}(\frac{\omega^2}{\omega^2+v_A^2q^2})\frac{\partial u'_z}{\partial z} -q^2(\frac{\omega^2 + v_{A}^{2} k_{x}^{2}}{\omega^2 + v_{A}^2 q^2})u'_z=0,
\end{equation}
where $q^2=\omega^2/c_s^2+k_x^2$, and, $v_A$ is Alfven speed, i.e. $v_A = B_0/\sqrt{4\pi\rho}$. This is an ordinary linear differential equation with constant coefficients and its general solution can be written as,
\begin{equation}\label{otin8}
 u'_z=A_1  \exp (s_+ z)+  A_2 \exp (s_- z),
\end{equation}
and the parameters $s_+$ and $s_-$ are obtained  as,
\begin{equation*}
s_{\pm}=\frac{g_{\rm eff}\omega^2}{2c_s^2(\omega^2+v_A^2q^2)}\pm
\end{equation*}
\begin{equation}\label{otin9}
\qquad\qquad\qquad\pm \sqrt{[\frac{g_{\rm eff}\omega^2}{2c_s^2(\omega^2+v_A^2q^2)}]^2+q^2(\frac{\omega^2 + v_{A}^{2} k_{x}^{2}}{\omega^2 + v_{A}^2 q^2})}.
\end{equation}

A general dispersion relation is obtained by imposing the following boundary conditions at the interface:  (1) The perturbations must tend to zero as $z$ goes to the infinity; (2) The $z$-component of the velocity is continuous at the interface; (3) The total pressure, including radiation, magnetic and gas pressures must be continuous at the interface. Using these boundary conditions and the fact that one of the roots is always positive and the other one is negative, we obtain $u'_z=A_2 \exp (s_- z)$ for the upper layer $(z > 0)$, and, $u'_z =A_1\exp (s_+ z)$ for the lower layer $(z < 0)$. Also, continuity of the total pressure at the interface is written as
\begin{equation}\label{otin10}
 (p'_g+p'_m-\rho g_{\rm eff}\zeta)\big\vert_{0^+}=(p'_g+p'_m-\rho g_{\rm eff}\zeta)\big\vert_{0^-},
\end{equation}
where $p'_m$ is the perturbed magnetic pressure, i.e. $p'_m=-(B_{0}^{2}/4\pi\omega)(\partial u'_{z}/\partial z)$ and  $ \zeta ={u'_z}/{\omega}$.  Dispersion relation, therefore, is obtained as 
\begin{equation*}
 \frac{\rho_2}{\omega^2+k_x^2c_{s_2}^2}\big(\omega^2 c_{s_2}^2 s_-+g_{\rm eff_2} k_x^2c_{s_2}^2\big) +{v_A}_2^2\rho_2 s_--
\end{equation*}
\begin{equation}\label{otin110}
\quad- \frac{\rho_1}{\omega^2+k_x^2c_{s_1}^2}\big(\omega^2 c_{s_1}^2 s_++g_{\rm eff_1} k_x^2c_{s_1}^2\big) -{v_A}_1^2 \rho_1 s_+=0,
\end{equation}
and this equation can be written as
\begin{equation*}
 \frac{1}{x^2+y^2\mu^2}\big(\alpha x^2S_-+\alpha(1-E_2) y^2\mu^2\big) -
\end{equation*}
\begin{equation}\label{otin11}
\quad- \frac{1}{x^2+y^2}\big(x^2S_+  +(1-E_1) y^2\big) +\beta_2^2\alpha S_--\beta_{1}^2 S_+=0.
\end{equation}
Here, the dimensionless parameters $x, y, S_{-}, S_{+}, \beta_1 $, and $\beta_2$ are defined as
\begin{equation}
 x=\frac{\omega}{g} c_{s_1},\quad y=\frac{k_x}{g}c_{s_1}^2,\quad\alpha=\frac{\rho _2}{\rho_1}, \quad \mu=\frac{c_{s_2}}{c_{s_1}},\nonumber
\end{equation}
\begin{equation}\label{otin12}
\beta_2=\frac{v_{A_2}}{c_{s_2}},\quad  \beta_1=\frac{v_{A_1}}{c_{s_1}}, \quad  S_{-}= \frac{s_{-}}{g}c_{s_2}^2,\quad S_{+}= \frac{s_{+}}{g}c_{s_1}^2.
\end{equation}
We note that not all the above defined dimensionless parameters are independent. In fact, pressure continuity implies that the parameters of $\alpha$ and $\mu$ are not independent and we have  $\alpha \mu^2=1 $. Furthermore,  continuity of magnetic pressure implies that $\beta_1=\beta_2$. Also, we have
\begin{equation*}
 S_-= \frac{(1-E_2)x^2}{2(x^2+\beta_2^2x^2+\beta_2^2\mu^2 y^2)}-\bigg[\big(\frac{(1-E_2)x^2}{2(x^2+\beta_2^2x^2+\beta_2^2\mu^2 y^2)}\big)^2
\end{equation*}
 \begin{equation}
\qquad\qquad+(x^2\mu^2+y^2\mu^4)(\frac{x^2 + \beta_{2}^{2}\mu^2 y^2}{x^2 + \beta_{2}^{2} x^2 + \beta_{2}^2 \mu^2 y^2})\bigg]^{1/2},
\end{equation}
\begin{equation*}
 S_+=(\frac{(1-E_1)x^2}{2(x^2+\beta_1^2x^2+\beta_1^2 y^2)}+\bigg[\big(\frac{(1-E_1)x^2}{2(x^2+\beta_1^2x^2+\beta_1^2 y^2)}\big)^2
 \end{equation*}
 \begin{equation}\label{otin14}
\qquad\qquad\qquad\qquad+(x^2+y^2)(\frac{x^2 + \beta_{1}^{2} y^2}{x^2 + \beta_{1}^{2} x^2 + \beta_{1}^2  y^2})\bigg]^{1/2}.
\end{equation}

Equation (\ref{otin11}) is our non-dimensional dispersion relation for analyzing the magnetic RT instability in an optically thin medium which can be solved numerically. Before presenting our numerical analysis of this equation, however, it is possible to obtain analytical solutions for a few simplified cases as we show in the next subsections.
\subsection{Dispersion relation in simplified cases}
If we ignore radiation and magnetic fields, we can set $E_1 = E_2 = 0$ and $\beta_1 = \beta_2 = 0$  into equation (\ref{otin11}), and thereby, the classical nonmagnetic dispersion relation is obtained which is the same as  equation (16) of \cite{shiva} for a similar configuration for $\mu=1$.   Moreover, the incompressible limit is obtained, if the sound speed in both the upper and the lower fluids tends to infinity. We then obtain
\begin{equation}\label{otin16}
 x^2 = \frac{\alpha-1}{\alpha+1}y,
\end{equation}
which is in agreement with the results of \cite{chandra} for a similar problem. It is found that the system is prone to the RT instability, if we have $\rho_2>\rho_1$, and, the compressibility does not change this criteria for the onset of  instability. As the wavelength of the perturbations becomes longer, however, reduction of the growth rate due to the compressibility becomes more noticeable. Furthermore, \cite{chandra} presented a linear analysis of the incompressible magnetic RT instability including a uniform magnetic field. It turns out a uniform tangential magnetic field slows down the growth rate of the RT instability. The growth rate for the unstable modes with wavenumber $k_x$ parallel to the magnetic field lines is given by the following equation
\begin{equation*}
\frac{1}{x^2+y^2\mu^2}\big(\alpha x^2S_-+\alpha y^2\mu^2\big)-
\end{equation*}
\begin{equation}\label{otin17}
\qquad\qquad- \frac{1}{x^2+y^2}\big(x^2S_+  + y^2\big) +\beta_2^2\alpha S_--\beta_1^2 S_+=0,
\end{equation}
and if the sound speed tends to infinity, the growth rate for the incompressible case with $\beta_1 = \beta_2 = \beta $ and $\mu =1$ is obtained, i.e.
\begin{equation}\label{otin18}
x^2=\frac{\alpha-1}{\alpha+1}y-\frac{2}{\alpha+1}\beta^2 y^2,
\end{equation}
which is consistent with the previous studies (e.g., Chandrasekhar 1961). 

Figure \ref{f2} shows growth rate of the unstable perturbations as a function of the wavenumber based on the roots of equation ($\ref{otin17}$)  for  $\mu=0.45$ and  $\alpha=5$ and different values of $\beta$ where $\beta_1 = \beta_2 = \beta $. The compressible and  incompressible cases are shown by solid and dashed lines, respectively. Magnetic field has a stabilizing role in the RT instability by reducing the growth rate in comparison to the nonmagnetic case. In the presence of magnetic field, maximum growth rate occurs for a particular wavenumber, whereas in the nonmagnetic case, the growth rate increases with increasing the wavenumber. The critical wavenumber corresponding to the fastest growth rate reduces as the magnetic field becomes stronger. Although  Figure \ref{f2} corresponds to a particular set of the input parameters, we also found a similar behavior for the other set of the input parameters.
\begin{figure}
\centerline{\includegraphics[width=10cm]{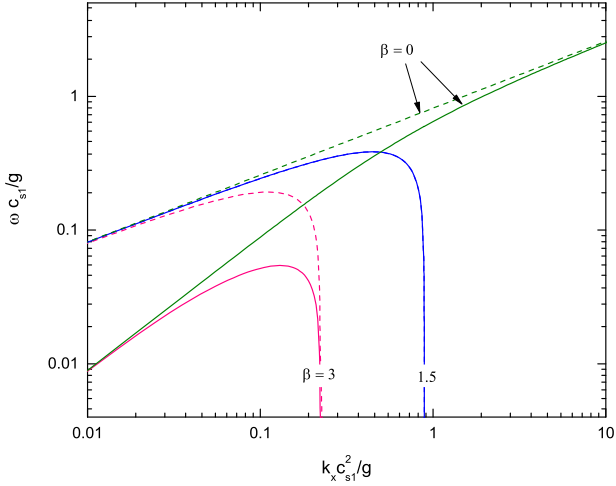}}
\caption{  Dimensionless growth rate  of the magnetic RT instability in the compressible case is obtained  by solving the full dispersion relation (\ref{otin17}) for $\mu=0.45$, $\alpha=5$, $\beta_1=\beta_2 = \beta$, and different values of $\beta$ (solid lines), however,  dashed lines show growth rate of the instability in the incompressible case.}
\label{f2}
\end{figure}
Role of the radiation field in RT instability, however, is explored  by setting $ \beta_1=0 $ and $ \beta_2=0 $ in equations (\ref{otin11}), (\ref{otin12}) and (\ref{otin14}). Then, the dispersion relation  can be written as
\begin{equation*}
\frac{1}{x^2+y^2\mu^2}\big(\alpha x^2S_-+\alpha(1-E_2) y^2\mu^2\big) -
\end{equation*}
\begin{equation}\label{otin19}
\qquad\qquad\qquad\qquad- \frac{1}{x^2+y^2}\big(x^2S_+  +(1-E_1) y^2\big) =0.
\end{equation}
This equation is the same as equation (74) of JK. In Figure \ref{f3} the dashed lines indicate unstable growth rates as a function of the wavenumber for $E_1=0$ and different values of $E_2$. The system is unstable when  the effective gravity is negative. In agreement with previous studies \citep[][JK]{jiang}, we also find that growth rate of the instability decreases with radiation pressure.  Figure \ref{f4} displays growth rate of the instability  for $E_2=0$ and different values of $E_1$. Here, the radiation force is assumed to be against the gravity (for having a unstable state). Therefore,  the input parameter $E_1$ is adopted  positive. \\
Up to now we have investigated radiative RT instability and  magnetic  RT instability separately. Using Equations (\ref{otin11}), (\ref{otin12}) and (\ref{otin14}), we can explore magnetic radiative RT instability in the optically thin regime. 
 Figure \ref{f3} shows growth rate of the RT instability including both magnetic and radiative effects as a function of the wavenumber of the perturbations for $\beta_1=\beta_2 =\beta = 3$, $\mu=0.45$, $\alpha=5$ and, $E_1=0$ (solid lines). In order to make easier comparison, we also show growth rate of the non-magnetic radiative RT instability by dashed lines. Each curve is labeled by the value of the corresponding Eddington parameter $E_2$. Moreover, dotted line represents a case with $E_1=0$ which tends to the classical RT instability. Figure \ref{f4} is similar to Figure \ref{f5}, but for $E_2=0$. This configuration may exist in boundary of a radiation-driven H II region. Growth rate of the RT instability reduces With increasing $\beta$. Magnetic fields strongly suppress the instability at long wavelengths, however, a radiation field reduces growth rate of the instability at all wavelengths. In the presence of the magnetic field, there is always a characteristic wavelength at which growth rate of the RT instability reaches to a maximum value. With increasing the Eddington limit of the background state $E$ which implies a stronger radiation field, the  wavelength of the fastest growing mode shifts to the longer wavelengths and the growth rate of the instability reduces. In other words, a radiation field in the optically thin regime has a stabilizing role on the magnetic RT instability. 
\begin{figure}
\centerline{\includegraphics[width=10cm]{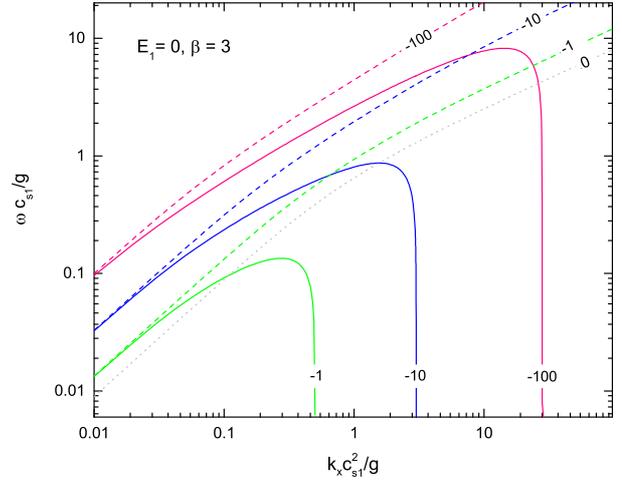}}
\caption{Growth rate of the magnetic radiative RT instability in the optically thin regime is shown by solid lines for $\beta =3$, $\mu=0.45$, and, $\alpha=5$. Each curve is labeled by the corresponding value of $E_2$. The dashed lines are radiative RT instability for the same parameters.}
\label{f3}
\end{figure}
\begin{figure}
\centerline{\includegraphics[width=10cm]{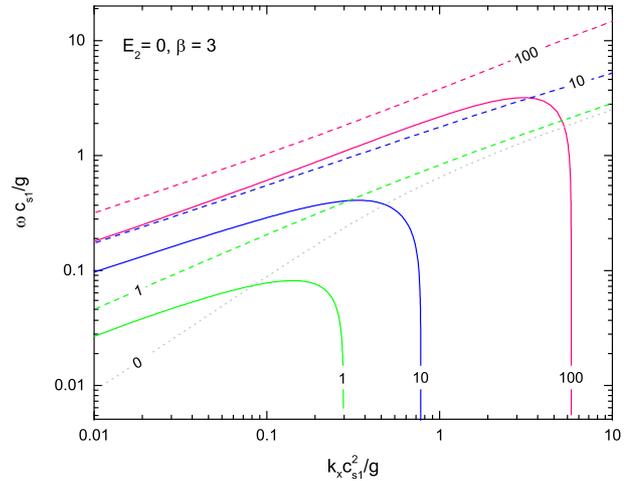}}
\caption{Same as Figure \ref{f3}, but for $E_2=0$ and each curve is labeled by the corresponding value of $E_1$.}
\label{f4}
\end{figure}
\section{ANALYSIS IN THE OPTICALLY THICK REGIME}
We now analyze RT instability  in the optically thick regime. Following JK approach in which the radiation field is assumed  to be Planckian with  $T_r = T_g =T$, and the comoving radiation pressure tensor to be isotropic, i.e. ${\cal P}_r = (E_r /3 ) {\cal I}_3$ and $E_r = aT^4$, we can simplify equation (\ref{eqs4})  as 
\begin{equation}
-\frac{\kappa_{\rm F} \rho}{c} {\bf F} \approx \nabla p_{\rm r},
\end{equation}
provided that  photon mean free path, i.e. $1/(\kappa_{\rm F}\rho)$, is smaller than the wavelength of the perturbations and the characteristic length scale of the system. Here, the Planck and energy opacities are assumed to be equal and constant.  Thus, equations (\ref{eqs2}) and (\ref{eqs3}) become
\begin{equation}\label{otic2}
\rho \frac{D{\bf u}}{Dt}=-\nabla p_{\rm t} + \rho {\bf g} + \frac{1}{4\pi} (\nabla\times {\bf B})\times {\bf B},
\end{equation}
\begin{equation}\label{otic3}
\frac{DE_{\rm t}}{Dt} + (E_{\rm t} + p_{\rm t} ) \nabla\cdot {\bf u} = -\nabla\cdot {\bf F}.
\end{equation}
where $p_{t}=p_g+p_r$.
 
\subsection{Initial equilibrium state}
As before we must specify the initial configuration which is supposed to be in thermal and mechanical equilibrium. As we now argue, however, the system can not fulfill  thermal and mechanical equilibrium conditions simultaneously unless some further assumptions are implemented. Since the initial state is in mechanical equilibrium and the upper and the lower layers have different densities, their temperatures can  not be the same. Thermal equilibrium, however, implies that both layers have the same temperature. In order to resolve this contradiction, JK proposed that one layer is optically thin and the second layer is optically thick. They also assumed that the system is adiabatic.   In the optically thin regime, where opacity on one or both sides of the interface is negligible, continuity of $E_r$ holds on whereas $T_g$ is independent of $E_r$. There are, however, further alternative ways to  specify background state self-consistently. \cite{jiang}, for instance, argued that even when the background state is not strictly in thermal equilibrium, its evolution affects subsequent thermal evolution very slightly so long as thermal time-scale is much  longer than the instability time-scale. Under these circumstances, therefore, the initial configuration can be imagined to be in thermal equilibrium for exploring RT instability.  \cite{jiang} constructed an initial state with zero  absorption opacity and effectively infinite thermal time-scale. As we mentioned earlier, JK assumed that  one side of interface is in the adiabatic regime, and the other side is in the optically thin regime. Here, we follow JK assumptions for specifying the initial equilibrium state which are applicable to some astrophysical systems.  \\

The momentum and the energy equations are written as
\begin{equation}\label{otic5}
 \frac{\partial p_{g}}{\partial z} =-\rho g(1-E),
\end{equation}
\begin{equation}\label{otic6}
 \frac{\partial E_{t}}{\partial t}=-\nabla.{\bf F}=0.
\end{equation}
where ${\bf F}=({-c}/{3\kappa_F\rho})\nabla E_{r}$. Equation (\ref{otic6}) is written based on the radiative equilibrium assumption. The parameter $E$ represents  the Eddington limit of the background state.  We note that the introduced parameter $E$ is then replaced by $E_1$ and $E_2$ for the lower and the upper layers, respectively.   Also, we have
\begin{equation}\label{otic7}
\frac{dT}{dz}=\frac{T}{4p_r}\frac{dp_{r}}{dz}=\frac{T}{4p_r}(-\rho gE).
\end{equation}
Introducing ratio of the radiation pressure to the gas pressure as  $r=p_{r}/p_{g}$, we can then write
\begin{equation}\label{otic8}
\frac{\partial \rho}{\partial z}=\frac{\rho g}{c_s^2}(E-1+\frac{E}{4r}),
\end{equation}
and
\begin{equation}\label{otic9}
\frac{\partial E_{t}}{\partial z} =\rho g(\frac{E-1}{\gamma-1}-3E).
\end{equation}
 We note that the Eddington limit parameter $E$ is a fixed input parameter, however, the parameter $r$ is not a constant under our imposed conditions. 

\subsection{Linear perturbations}
After deriving  the above equilibrium solutions, we can now  linearly perturb RMHD equations.  Following JK, we also apply {\it adiabatic approximation} by which we mean perturbed energy flux is neglected. It means that we require $F' \ll E_{\rm t} u'$. Using this approximation, we can assume that the opacity is a constant. Therefore, the parameter $E$ becomes a constant which is actually our implemented assumption. This parameter, however,  may be different in the upper and lower layers. We thereby arrive to the following equations:
\begin{equation}\label{otic10}
\omega \rho '+u'_z\frac{d\rho}{dz} +\rho (ik_x u'_x+\frac{\partial u'_z}{\partial z})=0,
\end{equation}
\begin{equation}\label{otic11}
\rho\omega u_z'=-\frac{\partial p_{t}'}{\partial z}-\rho' g-\frac{B_0}{4\pi}(\frac{\partial b'_x}{\partial z}-ik_x b'_z),
\end{equation}
\begin{equation}\label{otic12}
\rho\omega u'_y=\frac{B_0}{4\pi}(ik_x b'_y),
\end{equation}
\begin{equation}\label{otic13}
\rho\omega u_x'=-ik_x p_{t}',
\end{equation}
\begin{equation}\label{otic14}
\omega(\frac{1}{\gamma-1}p_g'+3p_r')+{\bf u'}.\nabla E_{t} -\frac{(E_{t} + p_{t})}{\rho} (\omega \rho '+u'_z\frac{d\rho}{dz})=0,
\end{equation}
\begin{equation}\label{otic15}
\frac{p'_g}{p_{g}}=\frac{\rho'}{\rho}+\frac{p'_r}{4p_{r}},
\end{equation}
\begin{equation}\label{otic16}
p'_{t}=p'_g+p'_r ,
\end{equation}
\begin{equation}\label{otic17}
\omega b'_x=-B_0(\frac{\partial u'_z}{\partial z}),
\end{equation}
\begin{equation}\label{otic18}
\omega b'_y=B_0(ik_x) u'_y ,
\end{equation}
\begin{equation}\label{otic19}
\omega b'_z =B_0(ik_x) u'_z ,
\end{equation}
\begin{equation}\label{otic20}
ik_x b'_x+ \frac{\partial b'_z}{\partial z}=0.
\end{equation}

As before, we assume all the perturbed quantities are proportional to $\exp(\omega t+ik_x x)$, and the above equations after lengthy but straightforward calculations reduce to the following equations:
\begin{equation}\label{otic21}
\omega \rho '+u'_z\frac{d\rho}{dz} +\rho(\frac{k_x^2}{\omega\rho} p'_{t}+\frac{\partial u'_z}{\partial z})=0,
\end{equation}
\begin{equation}\label{otic22}
\rho\omega u_z'=-\frac{\partial p_{t}'}{\partial z}-\rho' g +\frac{B_0^2}{4\pi\omega}(\frac{\partial^2 u'_z}{\partial z^2}-k^2_x u'_z),
\end{equation}
\begin{equation}\label{otic23}
 \frac{p'_{t}}{p_{g}}C=\frac{\rho'}{\rho}+\frac{gB}{c_s^2} \frac{u'_z}{\omega},
\end{equation}
where parameters $B$ and $C$ depend on the input parameters which specify the initial configuration of the system, i.e. 
\begin{equation}
B=\frac{1}{16r^2 +20r+\frac{\gamma}{\gamma-1}} [ 16r^2 (E-1)+r(24E-8)+
\end{equation}
\begin{equation} \label{otic25}
\qquad\qquad\qquad\qquad\qquad+E(5+\frac{\gamma}{\gamma-1})-1+\frac{E\gamma}{4r(\gamma -1)} ],
\end{equation}
%
\begin{equation}\label{otic24}
C=\frac{12r+\frac{1}{\gamma-1}}{16r^2+20r+\frac{\gamma}{\gamma-1}}.
\end{equation}

We now obtain $\rho'$ from equation (\ref{otic23}) and substitute it  into equations (\ref{otic21}) and (\ref{otic22}):
\begin{equation}\label{otic27}
 \frac{\partial u'_z}{\partial z}=\frac{g}{c_s^2}Cu'_z-\frac{1}{\rho \omega}(\frac{\omega^2C}{c_s^2}+k_x^2)p'_{t},
\end{equation}
\begin{equation}\label{otic28}
 \frac{\partial p_{t}'}{\partial z}= (\frac{\rho g^2}{c_s^2\omega} B -\rho\omega) u_z'- g\frac{C}{c_s^2}p'_{t}+\frac{\beta^2 \rho c_s^2}{\omega}(\frac{\partial^2 u'_z}{\partial^2 z}-k^2_x u'_z).
\end{equation}
 Substituting ${\partial u'_z}/{\partial z}$ from equation (\ref{otic27}) into equation (\ref{otic28}), this equation becomes
\begin{equation}
-\big(A\frac{\omega^2 c_s^2}{g^2}+\beta^2\frac{k_x^2 c_s^4}{g^2}\big)\frac{\partial p_{t}'}{\partial z}+\rho \omega\big(G-(\frac{\omega}{g}c_s)^2-\beta^2\frac{k_x^2}{g^2}c_s^4\big)u'_z-\nonumber
\end{equation}
\begin{equation}
-\frac{g}{c_s^2}\bigg[I \frac{\omega^2}{g^2}c_s^2-\beta^2(E-1+\frac{E}{4r}) \frac{k_x^2}{g^2}c_s^4\bigg]p_{t}'+\nonumber
\end{equation}
\begin{equation}\label{otic30}
\qquad\qquad\qquad\qquad\qquad \qquad\qquad+\beta^2\rho c_s^2C \frac{\omega}{g}\frac{\partial u'_z}{\partial z}=0,
\end{equation}
where
\begin{equation}\label{otic31}
A=\beta^2C+1, 
\end{equation}
\begin{equation}\label{otic33}
I=C-\beta^2(E-1)C-\beta^2H, 
\end{equation}
\begin{equation}\label{otic32}
 H=\frac{12-C(32r+20)}{16r^2 +20r+\frac{\gamma}{\gamma-1}} \big(E+r(E-1)\big),
\end{equation}
\begin{equation}\label{otic34}
 G=B+\beta^2(\frac{CE}{4r}-H).
\end{equation}
Note that the parameters $C$, $c_{\rm s}$, and, $r$ are not constant and their spatial dependence are obtained from the initial conditions, i.e. $\partial r/\partial z = -(g/c_{\rm s}^2 ) (E+x(E-1))$.

Now we consider perturbations which are actually localized at the interface. This means that we perform a WKB analysis of the vertical modes which it enables us to  eliminate the first order derivative terms. This condition is referred by JK as {\it adiabatic approximation}.
By assuming that quantities $p_{\rm t}'$ and $u_{\rm z}'$ are proportional to $\exp (s z)$, then equations  (\ref{otic27}) and (\ref{otic30}) become
\begin{equation}\label{otic36}
\frac{g}{c_s^2}\big( \frac{s}{g}c_s^2-C\big)u'_z+\frac{1}{\rho  \omega }\big(\frac{\omega ^2}{c_s^2}C+k_x^2\big)p'_{t}=0,
\end{equation}
and
\begin{equation}\label{otic37} 
\rho \omega \bigg(G-\frac{\omega^2 c_s^2}{g^2}-\beta^2 \frac{k_x^2 c_s^4}{g^2}+ \frac{\beta^2c_s^2}{g}Cs\bigg)u'_z-
\end{equation}
\begin{equation*}
-\frac{g}{c_s^2}\bigg[ \frac{c_s^2}{g}s\bigg(A\frac{\omega^2 c_s^2}{g^2}+\beta^2\frac{k_x^2 c_s^4}{g^2}\bigg)+I\frac{\omega^2 c_s^2}{g^2}-\beta^2Z\frac{k_x^2 c_s^4}{g^2}\bigg]p_{t}'=0,
\end{equation*}
where $Z=E-1+ E/4r$. From equations (\ref{otic36}) and (\ref{otic37}) by introducing $S= ({c_s^2}/{g})s $, we obtain
\begin{equation*}
 S^2(A \omega^2+\beta^2c_s^2k_x^2)-S(K \omega^2+\beta^2Zc_s^2k_x^2)-C(I\omega^2-\nonumber
 \end{equation*}
 \begin{equation}\label{otic39}
-\beta^2Zc_s^2k_x^2)+(G-\frac{\omega^2 c_s^2}{g^2}-\beta^2\frac{k_x^2c_s^4}{g^2})(\omega^2C+c_s^2k_x^2)=0,
\end{equation}
where $K=C-I$. Equation (\ref{otic39}) is solved numerically and a positive root, $S_{-}$, and a negative root, $S_{+}$, are obtained. 
Using these roots, the general solution can be written as equation (\ref{otin8}). Then, we apply boundary conditions. Continuity of the pressure at the interface is written as
\begin{equation}\label{otic42}
\bigg(p'_{t}+p'_m-\rho g\zeta\bigg)_{z=0^+}=\bigg(p'_{t}+p'_m-\rho g\zeta\bigg)_{z=0^-},
\end{equation}
where $p'_{t}$ is obtained from equation (\ref{otic27}) as follows 
\begin{equation}\label{otic43}
p'_{t} =\rho\omega\frac{C{g} u_{z}' /{c_s^2}-{\partial u'_z}/{\partial z}}{{\omega^2C}/{c_s^2}+k_x^2},
\end{equation}
and  $p'_m$ is given by
\begin{equation}\label{otic44}
p'_m=-\frac{B_0^2}{4\pi\omega}\frac{\partial u'_z}{\partial z}.
\end{equation}

Finally, the dispersion relation is obtained from equation (\ref{otic42}) and by imposing boundary conditions, i.e. 
\begin{equation*}
(S_-)(\frac{\alpha \omega^2}{C_2 \omega^2+c_{s_2}^2 k_x^2}+\alpha\beta_2^2)+ (\frac{\alpha c_{s_2}^2 k_x^2}{C_2 \omega^2+c_{s_2}^2 k_x^2})-
\end{equation*}

\begin{equation}\label{otic45}
\qquad\qquad-(S_+)(\frac{\omega^2}{C_1\omega^2+c_{s_1}^2 k_x^2}+\beta_1^2)- (\frac{ c_{s_1}^2 k_x^2}{\omega^2 C_1 + c_{s_1}^2 k_x^2})=0,
\end{equation}
where $S_-$ and $S_+$ are obtained from equation  (\ref{otic39}) for regions 2 and 1 , respectively. We note that in the absence of magnetic field (i.e., $\beta =0$), equation (\ref{otic45}) reduces to equation (74) of JK, i.e.

\begin{equation*}
 \frac{ \alpha}{(x^2C_2+y^2\mu^2)}(y^2\mu^2-\\
\end{equation*}
\begin{equation}
-x^2 \sqrt{C_2( C_2-B_2 )+x^2C_2\mu^2-\frac{y^2\mu^2}{x^2} B+y^2\mu^4})=\\
\end{equation}
\begin{equation}\label{otic48}
 =\frac{1}{(x^2C_1+y^2)}(y^2+x^2\sqrt{C_1( C_1-B_1 )+x^2C_1-\frac{y^2}{x^2}B+y^2} ).\nonumber
\end{equation}

Equation (\ref{otic45}) is a dispersion relation when two sides of interface are adiabatic. In addition to assuming  that one side (e.g., region 1) is in the optically thin regime, we also suppose that $\rho_2\gg \rho_1$ and $c_{s_2}\ll c_{s_1}$ and the magnetic field is not strong. Upon applying these simplifying assumptions, therefore, Equation (\ref{otic45}) reduces to 
\begin{equation}\label{otic49}
(S_-)(\frac{ \omega^2}{C_2 \omega^2+c_{s_2}^2 k_x^2}+\beta_2^2)+ (\frac{ c_{s_2}^2 k_x^2}{C_2 \omega^2+c_{s_2}^2 k_x^2})=0.
\end{equation}
 Equation (\ref{otic49}) is our main equation that we plan to investigate it in more detail.  In the long-wavelength limit, we have
\begin{equation}\label{otic50}
\omega^2=(\frac{g}{c_{s_2}})^2(G_2-I_2).
\end{equation}
If we set $\beta_2=0$,  JK relation (78) is recovered. The growth rate increases with magnetic field strength. A critical wavenumber is obtained by assuming that $\omega=0$. Thus, the critical wavenumber becomes  
\begin{equation}\label{otic51}
k_x^2=(\frac{g^2}{c_{s_2}^4})(C_2Z_2+\frac{G_2}{\beta_2^2}+\frac{Z_2}{\beta_2^2}+\frac{1}{\beta_2^4}).
\end{equation}
Perturbations with a wavelength longer than critical wavelength are completely suppressed in the presence of the magnetic fields. In the short-wavelength limit, however, the growth rate tends to equation (\ref{otic50}). 

Equation (\ref{otic49}) can be solved numerically for a given set of the input parameters.
Figure \ref{f5} shows the growth rates of the RT instability for different values of parameter $\beta_2$. Compressible RT instability is also shown by dotted line. Growth rate of the instability decreases because of considering magnetic fields.  We can also determine the fastest growing mode which is an important quantity. Before doing so, however, we discuss about validity of our results and the applied assumptions. 
\begin{figure}
\centerline{\includegraphics[width=10cm]{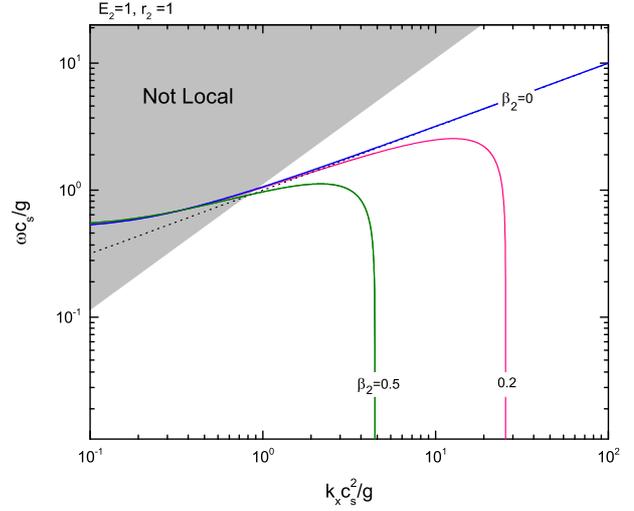}}
\caption{   Growth rate of the adiabatic, optically thick radiative RT instability in the presence of magnetic field. Solid lines represent cases with different parameter $\beta_2$. A case without magnetic field is also shown. Dotted line shows  growth rate  of the compressible RT instability. The gray region displays the evanescence constraint (the decay of perturbations on the length-scales smaller than the background equilibrium scale height) is violated according to equation (\ref{val3}).}
\label{f5}
\end{figure}
\subsection{Validity of the Approximations}
 We now discuss about validity of the imposed assumptions  in the optically thick regime. The requirement of the optically thick regime implies that $\tau={L}/{\lambda} \gg1$, where $L$ is the characteristic size of the system, i.e. $L={\rm min}(1+r,r/E)c_s^2/g$,  and $\lambda=1/\kappa_{\rm F}\rho$ is the photon mean free path. Thus, it must be smaller than the wavelength of the perturbations and the characteristic length scale of the system. Now we propose to consider evanescence condition that by which we have ${1}/{s}<L$, where $s$ is proportional to the perturbation in the  $z$ direction. Note that dimensionless parameter $s$ is defined in equation (\ref{otin12}). We can obtain $s$ from the equation (\ref{otic49}):
\begin{equation}\label{val1}
s_-=-g\frac{k_x^2}{\omega^2+\beta_2^2C_2 \omega^2+v_{A2}^2 k_x^2}.
\end{equation}
This term becomes negative, because the perturbations are finite as $z$ tends to the infinity in the positive direction of the $z$ axis. Now we can write the evanescence condition as:
\begin{equation}\label{val2}
\frac{\omega}{k_x}<c_s\sqrt{\frac{{\rm min}(1+r_2,r_2/E_2)-\beta_2^2}{1+\beta_2^2C_2}},
\end{equation}
and in the dimensionless form, it becomes
\begin{equation}\label{val3}
\frac{x}{y}<\sqrt{\frac{{\rm min}(1+r_2,r_2/E_2)-\beta^2_2}{1+\beta^2_2C_2}}.
\end{equation}

If we set $\beta_2 =0$, the above relation reduces to equation (76) of JK. The invalid region based on equation (\ref{val3}) is shown as a shaded area in Figure \ref{f5}  for $\beta_2 =0.5, 0.45$ and $E_2=r_2=1$.  The obtained results are valid over a wider range of the input parameters with increasing parameter $\beta_2$.  
As for the adiabatic approximation, we must have ${F_z}'\ll E_t {v_z}'$ to ignore of the perturbation of the radiation flux. Since we have $p_r'=-\xi (dp_r/dz)=E\rho g\xi$ and considering linearized  equation  of the radiation flux, we obtain
\begin{equation}\label{val4}
F'_z>\frac{-c}{\kappa \rho}\frac{dp'_r}{dz}=\frac{-c}{\kappa \rho} s p'_r=-sF_0\xi,
\end{equation}
and so,
\begin{equation}\label{val5}
\frac{F_0}{E_t}\ll-\frac{\omega}{s}=\frac{\omega}{gk_x}(\omega^2+v_A^2(\omega^2\frac{C_2}{c_s^2}+k_x^2)).
\end{equation}

This relation shows that the adiabatic approximation at long and small wavelengths is reliable. Considering  equations (\ref{val2}) and (\ref{val5}) for a given value of $F_0/E_{t}$, since the calculations do not depend on $F_0/E_t$, we can conclude that our results are valid. For strong magnetic fields, however, we note that the evanescence condition is violated.
\begin{figure}
\centerline{\includegraphics[width=10cm]{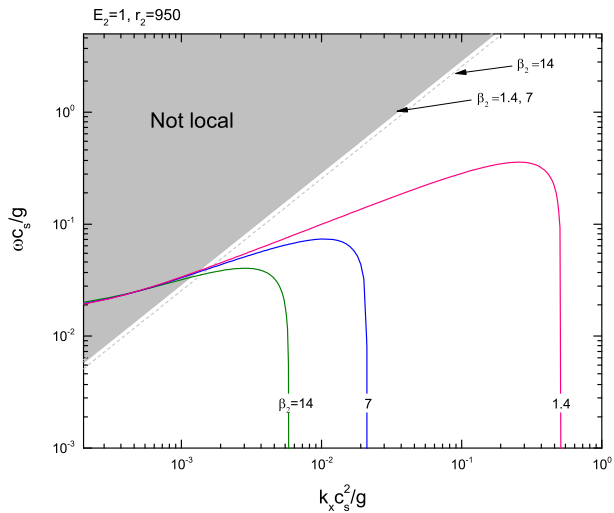}}
\caption{Same as Figure \ref{f5}, but for the parameters $E_2=1, r_2= 950$. }
\label{f6}
\end{figure}
\section{Astrophysical Implications: Radiative bubbles in the massive star forming regions}
We did a detailed analysis of the radiative RT instability in the presence of magnetic field. Our parameter study shows that magnetic fields are able to significantly modify unstable radiative RT modes which have already been studied by JK. We now present astrophysical implications of our study. JK  investigated radiative RT instability in the massive star forming regions without considering role of the magnetic field. Since radiation of a young star can be stronger than its gravitational force near the Eddington limit, further gas accretion onto the star can be prevented. Thus, the net force on the gas component is outward which pushes the gas toward outer regions. Under these circumstances, however, the interface between a low-density bubble and high-density shell that is generated from swept gas is prone to the RT instability. Even with a strong central radiation source, further gas accretion is possible as a result of non-linear growth of the RT instability \citep[][]{krum2009}. JK studied stability of this interface subject to RT instability in an extreme condition where the flow is near the Eddington limit. They found that linear growth time-scale of RT instability is less than 1000 years which is quite short compared to the star formation time-scale which is approximately of order $10^5$ years.

Observational evidences, however, indicate that magnetic fields may play a significant dynamical role in the photon bubbles of the star forming regions. Zeeman measurements show that the strength of the magnetic fields in these systems is about $0.01-0.6$ G  \citep{sarma2002,vlem}.  \cite{Turner2007} investigated radiative bubbles in the circumstellar envelopes of young massive stars with a magnetic field of around 0.1 G.

We now re-examine RT instability in the {\it magnetized} radiative bubbles. The central massive star is assumed to have a mass  $M_* = 100 $ M$_\odot$, and, its luminosity  is $L_* =10^5 $ L$_\odot$. So all the above parameters are calculated for the shell. At the edge of the bubble, we assume that physical quantities are $ \rho = 10^{-16}$ g cm$^{-3}$,  $T = 1100 K$, $\gamma=7/5$ (for molecular hydrogen), and, the mean mass per particle is $m = 2.33 m_{\rm H}$. Then, the  gas sound speed becomes $c_s\approx 2 $ km s$^{-1}$. In the dense gas, we assume that $ E_2 \approx 1 $ and $r_2=950$ (JK) and consider the strength of the magnetic field to be about a few tens milli-Gauss. Given all the input parameters, one can easily solve equation (\ref{otic49}) numerically. In Fig. \ref{f6}, we have $B_0=10, 50 , 100 $  mG which then imply that $\beta_2= 1.4, 7, 14$.  The minimum bubble-growth time, therefore, becomes about 180 kyr and 900 kyr for $\beta_2=1.4$ and 7, respectively. This time-scale is a few times longer than formation timescale. The corresponding wavelength, furthermore, is about $10^5$ AU which is roughly ten  times larger than bubble size. The allowed  region  according to equations (\ref{val2}) and (\ref{val5}) is shown as a shaded area in Fig. \ref{f6}. For $\beta_2= 14$ this area becomes slightly larger than a case with a lower $\beta_2$. So, we can verify that our results are reliable for this range of the magnetic field. If the central mass is assumed to be 10 solar masses, the growth time would be ten times longer than what we just obtained. Thus, growth rate of instability decreases, when a central star with a lower mass is considered.
Our stability analysis is a variant of the magnetic buoyancy instability in which magnetic field lines and the direction of the perturbations are in parallel. Importance of the magnetic buoyancy instability in the Galactic disc (including cosmic ray component) to explain formation of the molecular clouds has been emphasized by \cite{Parker66} and further developments towards understanding nonlinear evolution of this instability are based on the direct numerical simulations \citep[e.g.,][]{Matsu93,Basu97,Hanasz2003,Rodrigues16}. However, we did not explore magnetic buoyancy instability in the presence of a radiation field when the perturbations are perpendicular to the magnetic field vector. In Figure 6, for instance, the gas and magnetic pressures are comparable when magnetic field strength is $B_0 = 10$ mG, whereas magnetic pressure becomes much larger than the gas pressure for a case with $B_0 = 100$ mG.  It deserves further study to explore stability of this strongly magnetized configuration in the presence of a radiation field subject to the perturbations perpendicular to the magnetic field lines because the instability may become so efficient that gas can get to the star. 

\section{Conclusions}
In this study, we studied radiative and magnetic RT instability in the linear regime.  In the absence of radiation or magnetic field, our dispersion equation reduces to results of the previous studies (e.g., JK), however, we found new features of the instability when radiation and magnetic effects are significant. In the context of massive star formation,  radiation pressure of a young central star is able to prevent further accretion of mass unless some mechanisms like RT instability at the edge of the radiation-driven bubble around the star provide channels of mass accretion onto it \citep[e.g.,][]{rosen,kla}. Linear development of radiative RT instability at the edge of bubbles around massive stars is explored by JK who confirmed significant dynamical role of the radiation field in  this scenario of massive star formation. In JK and most of the previous numerical simulations of massive star formation, magnetic fields have been neglected for simplicity.  We found that magnetic field has a stabilizing role in the radiative RT instability. Although the present study is restricted to the linear perturbations, our results clearly demonstrate that trend of the radiative RT instability is significantly modified when magnetic effects are considered and in order to adequately describe the development of RT instability as a mechanism of creating channels of mass accretion onto young massive stars, this important physical ingredient can not be neglected.

There are also other astrophysical systems where our analysis is applicable. For instance, it has been proposed that  clumpy structures  in  the outflows  from supercritical accretion flows are created by the  radiative RT instability \citep[e.g.,][]{takeu,takeu14}. \cite{takeu} performed radiation hydrodynamic simulations for a supercritical accretion flows and found that clumps are formed due to RT instability with shapes more or less elongated along the outflow direction. They argued that since formation of clumps are observed in their non-magnetic simulations, mechanisms such as magnetic photon bubble instability is not needed for the clump formation. There are, however, strong theoretical evidences that magnetic fields play a vital role in the accretion flows. If RT instability is responsible for clump formation in these systems, our study shows that magnetic field is able to suppress the instability and not only the clumps may form over a longer period of time but also their size is dependent on the strength of the magnetic field. This issue deserves further investigations.

\section*{Acknowledgments}
We are grateful to Mark R. Krumholz for helpful comments and advices. We also thank the anonymous referee for a thoughtful report and constructive suggestions.

\bibliographystyle{mnras}
\bibliography{reference}

\bsp

\label{lastpage}
\end{document}